# IoT Forensic

## A digital investigation framework for IoT systems

Snehal Sathwara, Nitul Dutta
Computer Engineering Department
Marwadi University
Rajkot, Gujarat, India
ed.snehal@gmail.com; ann28dee@yahoo.co.in

Emil Pricop
Automation, Computers & Electronics Dept.
Petroleum-Gas University of Ploiesti
Ploiesti, Romania
e-mail: emil.pricop@upg-ploiesti.ro

*Abstract* –Security issues, threats, and attacks in relation with the IoT have been identified as promising and challenging area of research. Eventually, the need for a forensics methodology for investigating IoT-related crime is therefore essential. However, the IoT poses many challenges for forensics investigators. These include the wide range and variety of information, the unclear lines of differentiation between networks, for example private networks increasingly fading into public networks. Further, integration of a large number of objects in IoT forensic interest, along with the relevance of identified and collected devices makes forensic of IoT devices more complicated. The scope of this paper is to present a framework for IoT forensic. We aimed at the study and development of the link to support digital investigations of IoT devices and tackle emerging challenges in digital forensics. We emphasize on various steps for digital forensic with respect to IoT devices.

I. INTRODUCTION

The latest developments in the sensing capabilities and connectivity of the electronics devices led to the apparition of a very complex and challenging domain of Internet of Things – IoT. In this concept all the devices are interconnected between each another and also are connected to the Internet by using various standardized communication protocols. Often these IoT equipment can be remotely controlled in a very simple manner.

The interconnected devices are becoming more and more frequently the target of various attackers. IoT equipment using standardized protocols and commercial firmware and software have a large number of vulnerabilities that are exploited by cybercriminals and hackers.

Hackers obtain gradually experience and knowledge regarding the IoT devices vulnerabilities and exploitation. In the authors opinion this fact will allow the following period to become the "era of IoT hacking".

Usually, hackers will try to produce DDoS attack on the IoT devices that can cripple critical infrastructures, systems and even the normal way of life. Hackers practice exploitation on the IoT devices and use it as a gateway to deeper levels of a network where they gather private information, alter or delete the data or destroy the whole system [1]. Thingbots, RFID, Wearable, Smart Plug, Traffic Lights, Cameras, Automobiles, Airplanes, Digital Locks, Pacemaker, Rifle, Digital Weapon, Thermostats etc. are the most targeted spots where hackers can practice to compromise.

Hackers represent a real problem to any organization irrespective of the size and nature of their business. The attackers are aiming on various assets in order to interfere with or to block the daily operation of the target organization.

The main objective of the attackers is to deploy any possible technique to exploit available vulnerabilities in victim's electronic devices. The existing security measures are not 100% efficient, so the exploitation is still possible. The threats dynamic is very high, and the new vulnerabilities or new exploits are discovered daily, even hourly.

The most severe issue arises when the hackers are able to access sensitive information like credit card details, health information or system credentials. Using this data, the attackers can perpetrate identity theft or other cybercrimes, that are very complex to investigate compared to traditional criminal activities.

The complication arises due to the use of sophisticated electronic devices and techniques, which give the attacker the possibility of hiding the identity. The process becomes more complex when it interferes with IoT devices. The existence of this kind of equipment in a network architecture makes the attacker job easier. As a line of protection, the analysis of such crimes after occurrence is very important to in order to stop their recurrence and to develop adequate protection means. Digital evidence is the most important piece of a huge puzzle that helps the investigator to draw suitable conclusion about the suspected crime. Collecting correct and viable digital evidence ensures also the legal case admissibility, especially in cyber oriented crimes.

Network forensic methods, protocols and tools were already developed and are available for investigators. Unfortunately, these tools are not sufficient to perform a reliable investigation in IoT device, due to their complex and heterogenous structure. This is the reason why a considerable attention is required in this area in order to obtain an efficient and productive digital investigation process and to collect even deteriorated digital evidence with accuracy and timeliness.

It is believed that it is not very difficult to stumble on the potential evidence related to a crime in networked devices. The claim is vindicated due to existence of comprehensive network logs, various chat logs, predictable emails and social networking conversations. However, what make the investigation difficult in IoT are the proprietary data formats, protocols, and physical interfaces that come across the practice of evidence extraction. The IoT devices are comparatively more vulnerable and susceptible on networks because of immature security protocols available so far to protect from potential threats.

In order to perform a digital forensic investigation of an IoT device, it is essential to understand characteristics of this kind of devices along with their response to security breach. Furthermore, inclusion in the networked infrastructures of nearly infinite number of IoT device from various manufacturers makes digital investigation more challenging. Moreover, the data collected by IoT devices is personal and huge in volume and need timely analysis for proper identification of threats during the forensic procedure. Analyzing such a sensitive and voluminous data with timeliness is itself a challenging task.

In this paper, we present a framework for digital forensic of IOT devices in order to investigate a cybercrime on an IoT system. Our prime objective is to perform a comprehensive study on components of IoT devices to support digital investigations and tackle emerging challenges in digital forensics. We also aimed at figuring out the logical steps to perform a systematic investigation of compromised IoT device to find the evidence of attack and subsequent caused damages.

## II. RELATED WORKS

Although quite a lot work has been done in digital forensic, the volume of work done in IoT forensic is very limited until the date of writing this paper. There are few works found in cloud forensic and they have shown similarities with IoT digital forensic. In this section we are discussing few of the recent works carried out in IoT forensic.

Hegarty et al. [2] have discussed the challenges faced by digital forensic in IoT. The authors have proposed a deployment model for digital investigation in cloud computing environment. The work shows a general review and probable solutions along with a system framework. However, they have not suggested any implementation strategy of their proposal.

In [3], Malek Harbawi et al. have stated a model for IoT digital evidence acquisition. A theoretical framework for IoT forensic is depicted in their work. Various steps of investigation are shown in order to collect evidence for further investigation.

In reference [4], Edewede et al. have explained various challenges and approaches of IoT forensic. They have divided the IoT network in to three zones as internal network, middle layer and external network. They have also proposed a Next-Best-Thing Triage (NBT) Model for use in conjunction with the stated three zone approach. Their proposed model is claimed to serve as a beacon to incident responders. It also increases efficiency and effectiveness of IoT-related investigations.

Mehroush et al. [5] have discussed security enhancement of IoT in the terms of forensics. The authors have shown differences between traditional investigation and current forensics investigation scenario. There were found some restrictions for IoT forensics investigation for the smart devices. Also, the authors proposed mobility forensics trends based on the innovations characteristic to smart device. Their proposal emphasizes the need to use sensor-based activity and collected data. Their proposed model has not been implemented and tested in real environments.

MacDermott et al. [6] have explained challenges in IoT forensics for the Internet of Anything era. They have mentioned challenges in data acquisition (Logical and Physical), extraction and analysis of data grows in this IoT environment. The authors proposed a combination of cloud native forensics with client-side forensics (Forensics for companion devices) to study and develop a support system for the digital investigations and tackle emerging challenges in digital forensics. They proposed the development of digital forensic standards that can be used as part of overall IoT and IoA security and aid IoT based investigations for the long-term goal.

Mauro et al. [7] also discussed challenges and opportunities in the field of Internet of Things security and forensics. They have briefly discussed major security and forensics issues for the security and forensics challenges. They focused on privacy, security and forensics challenges in IoT environment.

## III. IoT ECOSYSTEM

An ecosystem represents a complex system that integrates interdependent components required to complete the working of the system under consideration [8]. It may comprise living and non-living things and objects that are interconnected and work together. So far to the best of our knowledge, there is no ecosystem exist for IoT forensic.

In this section of the paper, the authors are trying to identify various components that contributes to the inner working of IoT forensic, as presented in reference [9]. The presented ecosystem is however not complete and may need further investigation. Our prime intention is to identify and enumerate the various components and possible methods of evidence collection for investigating an IoT based crime.

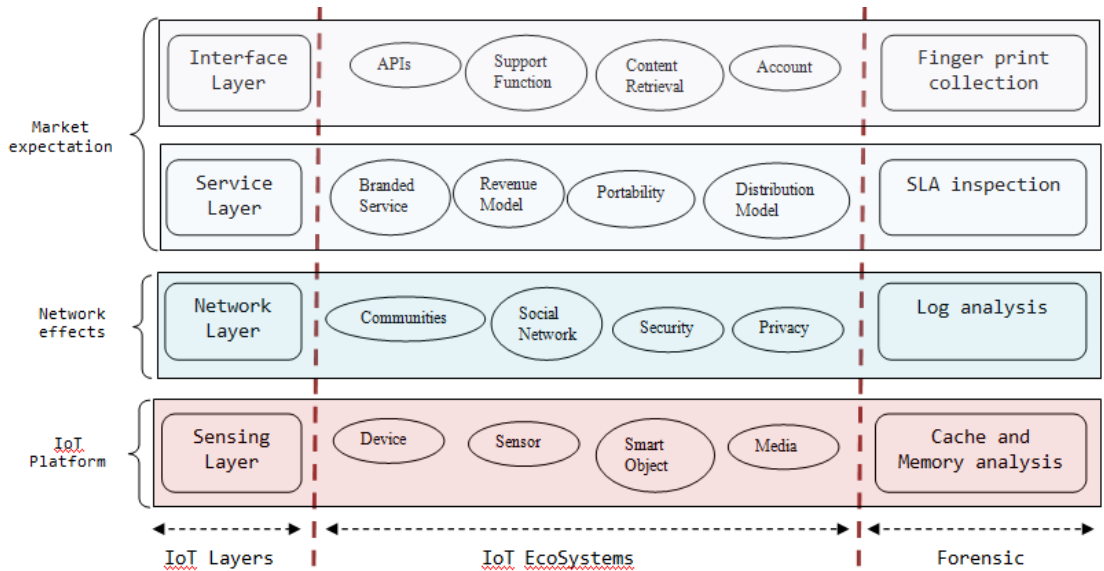
Figure 1 – IoT Forensic ecosystem

Figure 1 shows an ecosystem for IoT forensic. Three stacks are shown in parallel. In the first stack it is the layered architecture of IoT. The middle stack shows the components of IoT ecosystems and the third stack shows the possible forensic options.

Interface layer holds various APIs and content retrieval. For forensic of IoT various fingerprints in such applications can be analyzed. In the service layer revenue and distribution models are of critical importance. For this layer service level agreements [10] may be investigated to learn the behavior of an attack. The network logs from the communications made by the attacker may be analyzed as a part of IoT forensic. In the lower layer of the IoT protocol stack device level memories and caches may be examined for probable crime trace.

Interface layer and Service layer are considered as a part of market expectation, where application programming interface, support function, content retrieval and user credentials or accounts can be investigated by biometrics evidence collection.

Other side branded services, revenue generation model, portability options and distribution model can be investigated with the reference of service level agreement and continuous auditing or inspecting user and client agreement.

Network layer is considered is comprising all the network operations and their effect. Usage of internet, smart devices, data storage, data collection or retrieve, their security and privacy issues, potential usage of social media and their privacy concerns are initially investigated by collecting operation logs and analyzing those logs and collect suspicious evidence from them. It is also possible to collect time base information such as incident start time and end time. In the network layer it is critical to obtain the source and destination addresses and, if possible, malicious programming details.

Network type, routing information, attack vectors, details of existing firewall, details of existing network topology and equipment can be included also in the *Network effects layer* for the support of IoT ecosystem.

It is well established that sensors are the "heart", the real core, of any IoT system. This is the reason why the Sensing layer contains sensors, smart objects, smart devices and various connection media. In the IoT systems context the smart devices are playing a major role for making interconnections.

For the IoT Platform, respectively for the Sensing Layer, the investigation must focus on Cache and Memory analysis. Through memory analysis, there can be studied the limitation of transmission and the reasons for the attack executions. Collecting data and analyzing those data is big challenges to study and understand the IoT ecosystem.

## IV. IoT Forensic Steps

Figure 2 shows the steps that have to be carried out for digital forensic and subsequent challenges in IoT context. We are emphasizing on the end devices (sensors or smart equipment) roles and the challenges in analyzing their security. As presented in the previous sections of this paper forensic tools used for analyzing the classical computing or edge devices may be used in the case of IoT systems to a certain extent. Activity analysis is such devices is very difficult difficult as only limited information is stored, due to the limited computing resources in end devices

Search and seizure is an important step in any forensics investigation [11]. The most challenging part is identification of IoT or smart devices in the network or IoA environment because they are physically at nanoscale or tiny scale and passively automated.

In the case of IoT devices, which are part of larger networks, the evidence of cybercrimes is difficult to be gathered, due to the lack of tools and professional skills or to incorrect or insufficient documentation.

When an attack is happening, the hackers use all their skills to cover their tracks and to hide their own identity. In order to obtain some evidence of the criminal activity the forensic investigator should try to analyze the logs, which are playing a major role in this process. Also, a global view of the target system, attack mechanism and attacker possible motivation is needed.

The biggest challenge for the forensics in Internet of Things environments is represented by the preservation of whole digital crime scene [12]. It is not an easier task to collect data from a highly dynamic environment that concentrates heterogenous hardware and software architectures and variable resources (computing power, memory, storage space). In some devices all the logs are deleted when the device is powered off, making the forensic investigator task almost impossible.

Lack of proper tools to set up A to Z crime scene and to preserve information that collected from the sensors are another highly challenging area. The huge amount of data, including sometimes unnecessary data, is also interrupting the capacity of preservation. Now hackers are using most effective skills and tools to cover their identity. Due to this log are not getting up to date or showing fake route or identity. Most of IoT nodes are not storing any metadata including temporal information which makes origin of evidence a challenge for an investigator. In the case of modified data or temporal data are missed then correlation of evidences gathered from the different IoT devices is almost impossible task [7].

Finally, if all above three forensics steps are succeeding to gather and collecting evidence then performing presentation of the crime scene with proper evidence will be easy and fruitful.

## V. CONCLUSION

A framework for IoT forensic is presented in this paper. The authors introduced the forensic ecosystem that helps investigators in information gathering process. The steps for forensic gathering were identified along with probable challenges in acquiring evidence from crime scene. This work is an overview of the forensic investigation procedure and it should help in producing meaningful evidence in IoT crime.

Future research should focus on creating a framework for IoT forensic based on correlation of data and metadata from IoT nodes in order to overcome the challenges presented in this paper.

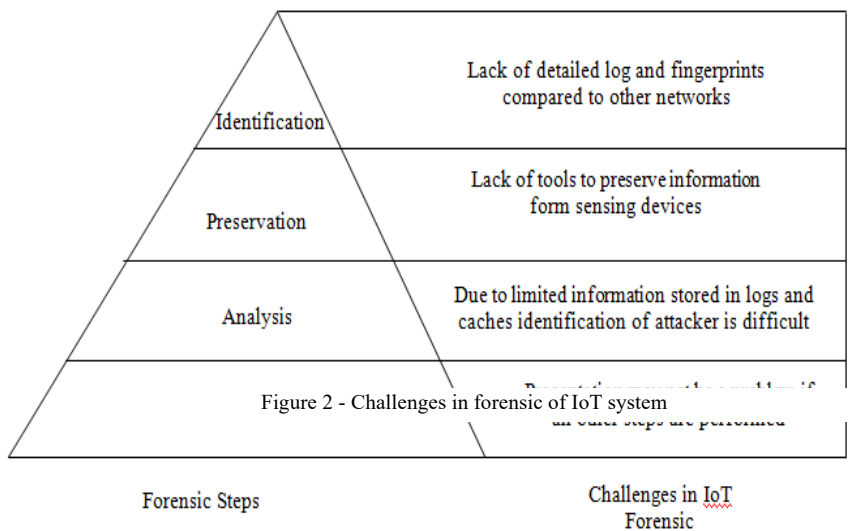

Figure 2 - Challenges in forensic of IoT system

## REFERENCES

[1] M. Dlamini, M. Eloff, "Internet of Things: Emerging and future scenarios from an information security perspective", SATNAC Intl. Conf. Proceedings, Swaziland, 2009.

[2] R. Hegarty, D.Lamb A Attwood, "Digital evidence challenges in the internet of things" Proc. of the 9th Int. Workshop on Digital Forensics and Incident Analysis, 2013, pp. 163-172.

[3] M. Harbawi, A. Varol, "The role of digital forensics in combating cybercrimes," 2016 4th International Symposium on Digital Forensic and Security (ISDFS), 2016, pp. 138-142.

[4] E. Oriwoh, D. Jazani, G. Epiphaniou, P. Sant, "Internet of Things Forensics: Challenges and approaches," 9th IEEE International Conference on Collaborative Computing: Networking, Applications, 2013, pp. 608-615.

[5] M. Banday, "Enhancing the security of IOT in forensics," 2017 International Conference on Computing and Communication Technologies for Smart Nation (IC3TSN), Gurgaon, 2017, pp. 193-198.

[6] A. MacDermott, T. Baker, Q. Shi, "Iot Forensics: Challenges for the Ioa Era," 2018 9th IFIP International Conference on New Technologies, Mobility and Security (NTMS), Paris, 2018, pp. 1-5.

[7] M. Conti, A. Dehghantanha, K. Franke, S. Watson "Internet of Things security and forensics: Challenges and opportunities" Editorial in Future Generation Computer Systems, 78 (2018) pp 544 – 546

[8] S. Zawoad, R. Hasan, "FAIoT: Towards Building a Forensics Aware Eco System for the Internet of Things," 2015 IEEE International Conference on Services Computing, New York, NY, 2015, pp. 279-284.

[9] P. Sundresan, N. M. Norwawi, V. Raman, " Internet of Things (IoT) digital forensic investigation model: Top Down forensic approach methodology," 5th Intl. Conf.e on Digital Information Processing and Communications (ICDIPC), 2015, Sierre, pp.19-23

[10] E. Fleisch, "What is the Internet of Things? An economic perspective." Economics, Management and Financial Market, 2010.

[11] J.Liu, "IoT Forensics: Issues and Challenges", at 12th IDF Annual Conference, 15th December 2015.

[12] S. Khan, "The Role of Forensics in the Internet of Things: Motivations and requirements", in IEEE Internet Initiative eNewsletter, July 2017